# (Newtonian) Space-Time Algebra

**J. E. Smith**

**University of Wisconsin-Madison (Emeritus)**

**Carnegie Mellon University (Adjunct)**

**02/14/2020**

## Abstract

The *space-time* (*s-t*) *algebra* supports a mathematical model for communication and computation that operate on values encoded as events in discretized linear (Newtonian) time. Consequently, the input-output behavior of *s-t* algebra and implemented functions are consistent with the flow of time. The *s-t* algebra and functions are formally defined. A network design framework for *s-t* functions is described, and the design of temporal neural networks is discussed as an extended case study. Finally, the relationship with Allen's interval algebra is briefly discussed.

## 1. Introduction

The *space-time* (*s-t*) *algebra* supports a mathematical model for communication and computation using values encoded as events in discretized linear (Newtonian) time. Consequently, the input-output behavior of implemented functions is consistent with the flow of time. Once formalized, the functional capabilities of the *s-t* algebra, and therefore the capabilities of temporal computing, can be fully explored. Most importantly, this includes the development of computing devices based on temporal principles.

The *s-t* algebra contrasts with conventional temporal algebras that are targeted at reasoning about temporal relationships, e.g., *interval algebras* [1]. These algebras essentially convert time intervals into a spatial form which facilitates the analysis of temporal relationships. This approach leads to algorithms that are aligned with the spatial manner with which we (humans) tend to reason about time [3]. In contrast to interval algebras, *s-t* algebra is an *event* algebra. Section 6 contains further discussion of Allen's algebra and its relationship to *s-t* algebra.

The *s-t* algebra embodies a temporal computing paradigm, as is hypothesized for biological neurons in the neocortex [8][11]. In biological neurons, values are encoded as voltage *pulses* or *spikes*. The occurrence of a spike is an event, and its relative time with respect to the times of other events encodes a value. Because of their transient temporal nature, spike-based computing paradigms can be modeled in a natural way with the *s-t* algebra.

For the construction of temporal computing machines, the role of *s-t* algebra is analogous to the role of Boolean algebra in conventional digital hardware.

## 2. Space-Time Algebra and Functions

Refer to Figure 1.

**Definition:** The *s-t algebra* is a bounded distributive lattice $\mathbb{S} = ( \mathrm{N}_0^\infty, <, \wedge, \vee, 0, \infty)$. $\mathbb{S}$ consists of a bottom element 0, a top element ∞, and the natural numbers. $\mathbb{S}$ is well-ordered and is closed under addition. □

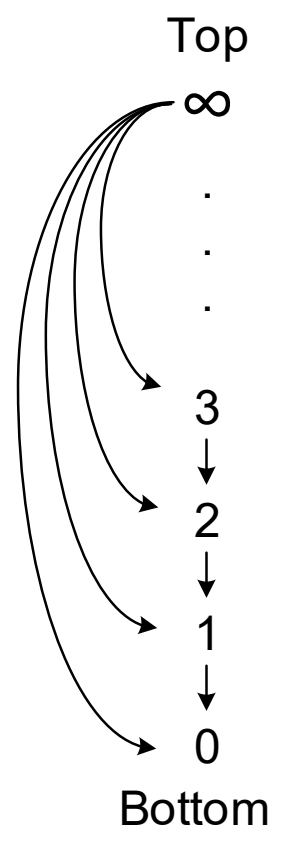

**Figure 1. The *s-t* algebra is a bounded distributive lattice.**

The top element is represented with the symbol "∞", not to be confused with the mathematical infinity. This symbol is chosen because the top element of the lattice has characteristics one would readily associate with an intuitive "∞". E.g., $\infty = \infty + 1$. It is useful for representing the case where a given event never occurs. That is, if an event occurs, it is given a value (time) that is relative to the occurrence of other events; if an event never occurs, it is given a value of ∞.

By definition, $\wedge$ and $\vee$ are distributive, associative, commutative, and satisfy the absorption laws. However, $\mathbb{S}$ is not complemented. Furthermore, there is no subtraction because the algebra is closed only for addition, and all elements of the algebra are non-negative. When the algebra is given a temporal interpretation then subtraction (or complementation or negation) would be tantamount to going backwards in time. Hence, such operations are not supported in the algebra.

As stated in the introduction, the algebra is intended to support physically implemented computing devices that communicate by encoding values as temporal events, for example values communicated as the times of electrical transients. In some implementations transient events may be voltage pulses or spikes transmitted over wires, or they may be changes in voltage levels (i.e., *edges*). In general, any method that relies on communication via transient temporal events may be used; photonic pulses communicated through free space is another example. In the remainder of this document, the terms "spike" and "pulse" are synonymous, and both are used as a generic way of denoting a point in time.

Of interest here is a certain class of functions defined over the *s-t* algebra. These functions take spikes as inputs, produce spikes as outputs, and the functional relationships between input spike times and output spike times are consistent with rules governing the flow of Newtonian time.

**Definition:** A function $z = F(x_1...x_q)$, $x_{1...}x_q$ , $z \in N_0^\infty$ , is a *Space-Time Function* if it satisfies the following:

1) *implementability:* $q$ is finite, and F is implementable with a finite number of states.

2) *causality*: i) For all $x_j > z$, $F(x_1...,x_j,...x_q) = F(x_1...,\infty,...x_q)$, and ii) if $z \neq \infty$, then $z \geq min(x_i)$ .

3) *invariance*: $F(x_1 + 1, ..., x_q + 1) = F(x_1,..x_q) + 1$. □

The three properties are very general. *Implementability* restricts consideration to functions that satisfy Church-Turing computability constraints. The other two properties are consistent with the uniform passage of time. *Causality*: using the temporal interpretation, an output spike can not be affected by input spikes that occur later in time. Furthermore, there are no spontaneous output spikes. *Invariance*: if all the input spikes uniformly shift by unit time, then the output spike shifts by unit time. Invariance naturally extends to any constant number of unit time shifts.

### 2.1 Primitive Space-Time Functions

In this section a set of *s-t* primitive functions or *operators* are defined in terms of the lattice's ordering relation $<$ .

#### 2.1.1 Unary Operators

The identity function, $a = a$, is one of two unary operators. The *increment* function, $a = b + 1$, is the other.

***Defn:*** For $b \neq \infty$, $a = b + 1$ if $b < a$ and there exists no $c < a$ such that $b < c$.
For $b = \infty$, $\infty = \infty + 1$.

The increment function naturally extends to the addition of any constant.

#### 2.1.2 2-ary Operators

Ordering relationships are a natural way of describing the 2-ary[1] operators. Table 1 captures all such functions of two inputs. The three left columns are associated with a set of three disjoint ordering relationships between inputs $a$ and $b$: $a < b$, $a = b$, and $b < a$. For a given operation and for each of these three input relationships, there are three possible outputs: $a$, $b$, or $\infty$. This suggests $3^3 = 27$ total 2-input functions. However, after accounting for symmetries and removing duplicates, there are 10 unique 2-ary operations; all are shown in Table 1.

**Commutative 2-ary operations:** *min, max, x-min, x-max, equals*

**Non-commutative 2-ary relations:** *not equals, less than, less or equals, greater, greater or equals*

In all cases, if the stated relation is true, then the output is equal to one of the inputs (the first input, $a$, by convention here); if the relation is false, then the output is $\infty$. Hence, for example, $a \prec b$ and $b \succ a$ are not the same function.

[1] Avoiding the term "binary" because of ambiguity with respect to binary encoded data.

**Table 1. All 2-ary *s-t* functions.**

| $a < b$ | $a = b$ | $b < a$ | function | name | symbol |
|---|---|---|---|---|---|
| $a$ | $a$ or $b$ | $b$ | if $a < b$ then $a$ ; else $b$ | *min* | $\wedge$ |
| $a$ | $a$ or $b$ | $\infty$ | if $a \leq b$ then $a$ ; else $\infty$ | *less or equal* | $\preccurlyeq$ |
| $a$ | $\infty$ | $a$ | if $a \neq b$ then $a$ ; else $\infty$ | *not equal* | $\neq$ |
| $a$ | $\infty$ | $b$ | if $a < b$ then $a$ else if $b < a$ then $b$ ; else $\infty$ | *exclusive min* | x$\wedge$ |
| $a$ | $\infty$ | $\infty$ | if $a < b$ then $a$ ; else $\infty$ | *less than* | $\prec$ |
| $b$ | $a$ or $b$ | $a$ | if $a \geq b$ then $a$ ; else $b$ | *max* | $\vee$ |
| $b$ | $\infty$ | $a$ | if $a > b$ then $a$ else if $b > a$ then $b$ ; else $\infty$ | *exclusive max* | x$\vee$ |
| $\infty$ | $a$ or $b$ | $a$ | if $a \geq b$ then $a$ ; else $\infty$ | *greater or equal* | $\succcurlyeq$ |
| $\infty$ | $a$ or $b$ | $\infty$ | if $a = b$ then $a$ ; else $\infty$ | *equal* | $\equiv$ |
| $\infty$ | $\infty$ | $a$ | if $a > b$ then $a$ ; else $\infty$ | *greater than* | $\succ$ |

## 2.2 Symbols and Notation

Symbols for the primitive functions are shown in Figure 2.

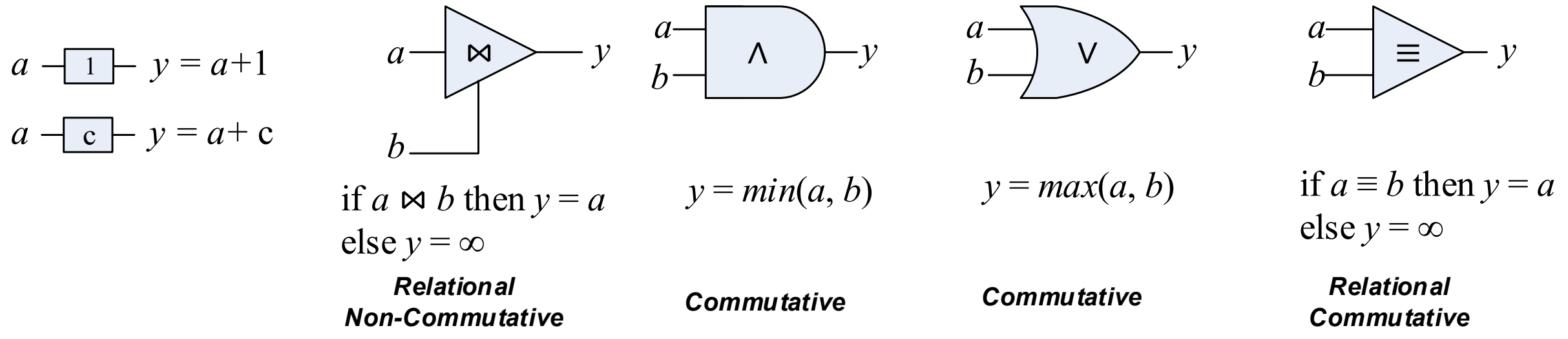

**Figure 2. Symbols representing the various primitive operators that may be used in network schematics. The symbol ⋈ represents any of the non-commutative relational operations.**

In written notation, the asymmetry of inputs is represented by the order of the inputs; that is, $a \bowtie b$ is always interpreted as: "if $a \bowtie b$ then $a$; else $\infty$". The input asymmetry in the drawn relational symbols mnemonically express their non-commutative relationship.

## 3. Basic Identities and Theorems

As noted above, the $\wedge$ and $\vee$ are commutative, associative, distributive, and satisfy the absorption laws. When the relations and +1 function are added, other useful identities arise.

***Basic Properties***

| | |
|---|---|
| ***1:*** $a \vee \infty = \infty$ | ***lattice top/bottom*** |
| ***2:*** $a \wedge 0 = 0$ | |
| ***3:*** $a \vee 0 = a$ | ***identity*** |
| ***4:*** $a \wedge \infty = a$ | |
| ***5:*** $a = a \wedge a$ | ***idempotence*** |
| ***6:*** $a = a \vee a$ | |
| ***7:*** $a \vee b = b \vee a$ | ***commutative*** |
| ***8:*** $= b \wedge a$ | |
| ***9:*** $a \vee (b \vee c) = (a \vee b) \vee c$ | ***associative*** |
| ***10:*** $a \wedge (b \wedge c) = (a \wedge b) \wedge c$ | |
| ***11:*** $a \wedge (b \vee c) = (a \wedge b) \vee (a \wedge c)$ | ***distributive*** |
| ***12:*** $a \vee (b \wedge c) = (a \vee b) \wedge (a \vee c)$ | |
| ***13:*** $a \wedge (a \vee b) = a$ | ***absorption*** |
| ***14:*** $a \vee (a \wedge b) = a$ | |
| ***15:*** $a \wedge a+1 = a$ | |
| ***16:*** $a \vee a+1 = a+1$ | |
| ***17:*** $(a \vee b) +1 = (a +1) \vee (b +1)$ | ***invariance*** |
| ***18:*** $(a \wedge b) +1 = (a +1) \wedge (b +1)$ | |
| ***19:*** $(a \prec b) + 1 = (a +1) \prec (b +1)$ | |

***Primitive Completeness; start with $\prec$, $\wedge$***

*20*: $a \succcurlyeq b = a \prec (a \prec b)$

***21:*** $a \preccurlyeq b = a \prec (b \prec a)$

***22:*** $a \vee b = (a \succcurlyeq b) \wedge (b \succcurlyeq a)$

***23:*** $a \succ b = (b \prec a) \vee a$

***24:*** $a \equiv b = (a \preccurlyeq b) \vee (b \preccurlyeq a)$

***25:*** $a \neq b = (a \prec b) \wedge (a \succ b)$

***26:*** $a \mathbf{x}\wedge b = (a \prec b) \wedge (b \prec a)$

***27:*** $a \mathbf{x}\vee b = (a \succ b) \wedge (b \succ a)$

***Some useful identities***

***28:*** $a \succcurlyeq b = (b \preccurlyeq a) \vee a$

***29:*** $a \preccurlyeq b = (a \prec b) \wedge (a \equiv b)$

***30:*** $a \succcurlyeq b = (a \succ b) \wedge (a \equiv b)$

***Distributive: combinations involving relationals***

***31:*** $a \prec (b \wedge c) = (a \prec b) \vee (a \prec c)$

***32:*** $a \prec (b \vee c) = (a \prec b) \wedge (a \prec c)$

***33:*** $(a \wedge b) \prec c = (a \prec c) \wedge (b \prec c)$

***34:*** $(a \vee b) \prec c = (a \prec c) \vee (b \prec c)$

***35:*** $(a \equiv b) \prec c = (a \prec c) \wedge (a \equiv b)$

***36:*** $a \prec (b \equiv c) = (a \prec b) \wedge (b \equiv c)$
***37:*** $(a \equiv b) \equiv c = (a \equiv c) \wedge (b \equiv c)$
***38:*** $a \equiv (b \wedge c) = [(a \equiv b) \vee (b \prec c)] \wedge [(a \equiv c) \vee (c \prec b)]$
***39:*** $a \equiv (b \vee c) = [(a \equiv b) \vee (c \prec b)] \wedge [(a \equiv c) \vee (b \prec c)]$
***40:*** $(a \prec b) \prec c = (a \prec b) \vee (a \prec c)$
***41:*** $a \prec (b \prec c) = (a \prec b) \wedge [(b \succcurlyeq c) \vee a)]$

$= [(a \prec b) \wedge [(c \preccurlyeq b) \vee b]] \vee a$

$= [(a \prec b) \wedge (c \preccurlyeq b)] \vee (a \wedge b) \vee a$

$= [(a \prec b) \wedge (c \prec b) \wedge (c \equiv b)] \vee a$

***42:*** $(a \bowtie_1 b) \bowtie_2 c = a \bowtie_1 b \ \vee \ a \bowtie_2 c$ ***meta-theorem; holds for any relational operators***

***e.g.*** $(a \prec b) \prec c = (a \prec b) \vee (a \prec c)$

## 4. Space-Time Computing Networks

*S-t* computing networks can be constructed by composing the 1-ary and 2-ary *s-t* operators. This is analogous to the construction of binary (Boolean) networks by composing *and*, *or*, and *not* gates.

**Lemma 1:** Any function implemented as a non-recurrent composition of space-time functions is a space-time function.

**Proof outline**: A proof begins with a topological sort of the directed graph implied by the non-recurrent composition. Then, the proof proceeds by induction on the sequence of elements as they occur in the topological sort. □

**Definition:** A *Space-Time Computing Network is* a non-recurrent (feedforward) interconnection of space-time functional blocks. Each block represents an implementable, causal, and invariant function. □

From Lemma 1, all space-time computing networks implement space-time functions, so if the designer begins with a set of basic operators that implement space-time functions and the operators are interconnected in any feedforward manner, the overall system must implement a space-time function.

### 4.1 Case Study: Boolean Design Framework

The Boolean design framework provides a good roadmap for developing a space-time design framework. It consists of several inter-related components that collectively form the basis of digital design and analysis. Working within the basic framework, logic designers *represent* Boolean function implementations and CAD tools *manipulate* these representations to improve the optimality of a design.

The following paragraphs describe important components of the Boolean-based framework. The components are described below in a sequence intended to emphasize their relationships.

***Arbitrary Form***

Boolean functions can be represented in *arbitrary form* as nested algebraic expressions composed of 1-ary and 2-ary operators (or "gates" using network terminology). These arbitrary multi-level expressions map directly to multi-level logic networks. In practice, a wide variety of operators are used: *and*, *or*, *not, nand, nor, xor,* etc. Because of the close relationship between algebraic expressions and networks, a designer can represent a function either algebraically or by drawing a schematic network representation.

***Standard Form***

A commonly used Boolean algebra *standard form* consists of fanned-out primary inputs, some of which may feed *not* operators, followed by a level of *and* operators, feeding into a single *or* operator which yields the primary output[2]. This is the standard sum-of-products form. There is also a product-of-sums standard form based on the duality of Boolean algebra. Because sum-of-products is more commonly used in practice, this discussion focuses only on the sum-of-products form. Any arbitrary feedforward network can be converted to sum-of-products form by following these steps:

1) Remove any internal fanout by replicating fanned-out subnetworks so that all network fanout is at the primary inputs.

2) Convert all logic gates to their *and*, *or*, and *not* equivalents.

3) Repeatedly apply deMorgan's law to push all *not* gates to the primary inputs; collapse series of *not* gates using double negation ($\neg \neg a = a$)

4) The network is now free of internal *not* gates. Start at the network output and repeatedly apply distributive, associative, commutative properties to reduce the network to sum-of-products standard form.

***Canonical Form***

The *canonical* form is a special-case standard form. In the (minterm) canonical form, every *and* operator (implicant) takes all the primary inputs, either with or without a *not*.

Any standard form network can be expanded to canonical form. The key theorem for doing so is:
$a = a \neg b + ab$.

The canonical form is uniquely defined for each function, and every standard form representation (and therefore any arbitrary representation) can be reduced to minterm canonical form. A given function may have many sum-of-products implementations, but only one canonical implementation.

***Truth Table***

A truth table is a representation that maps directly to the minterm canonical form; i.e., minterms map 1-to-1 with truth table rows. All truth table entries are either 0 or 1. By taking the *or* of all minterms with truth table output = 1, the minterm canonical form is obtained. Every function can be reduced to its unique canonical form. Completeness of Boolean algebra follows from the existence of a canonical form for every function (truth table).

***Implicant Table or Extended Truth Table***

In the *implicant table* there is one row per implicant for a standard form implementation. In addition to 0 and 1 entries, an implicant table typically contains the special symbol "-" to indicate an input that can be either 0 or 1. A given function may have many equivalent implicant tables (with the truth table being one of them).

***Discussion***

All the above components are important parts of an overall logic design framework. There are multiple algebraic representations (arbitrary, standard, canonical) and tabular representations (truth table, implicant table). Collectively, these representations are useful for specifying functions, for understanding functions, and for constructing and optimizing network implementations. Transformations between the representations are well-understood and widely used. For example:

1) A designer can specify a function as a multilevel network containing a variety of gate types. This can be done via a hardware design language or schematic diagrams, for example.

2) A designer can specify a function as an implicant table or truth table. Or a CAD tool may maintain equivalent data structures.

3) A function in arbitrary form or standard form can be reduced to minterm canonical form.

[2] This assumes that the 2-ary associative operators *and* and *or* generalize to *n*-ary equivalents.

4) A function in canonical form can be transformed into a truth table.

5) A function described as a truth table or implicant table can provide input to CAD optimization algorithms.

***Operator Choice***

When we design digital logic, we tend to use a set of operators that are intuitively appealing, regardless of the underlying technology -- i.e., we typically use *and*, *or*, *not*. These are also the operators that underly the standard and canonical forms.

In contrast, when we physically implement digital logic, we use operators that are best suited to the hardware technology. For example, in many digital circuits technologies these are typically *nand, nor,* or *aoi* gates.

That is, there is a set of *design* operators and a possibly different set of *implementation* operators. If they are different sets, then mapping algorithms can easily translate a design into an implementation. In the mapping process some optimizations can be performed.

*In summary: The objective is to develop a s-t design framework with capabilities and flexibility that are similar to the capabilities and flexibility of the Boolean design framework.*

## 4.2 Primitive Completeness

All the 2-ary operations can be constructed from only three primitives +1, $\wedge$, and $\prec$. In other words, these three primitives are *functionally complete* with respect to all the 2-ary functions. For example, Figure 3 shows constructions of $\succcurlyeq$, $\preccurlyeq$, $\vee$, and $\equiv$ .

***cases: ( a < b , a = b, b < a )***

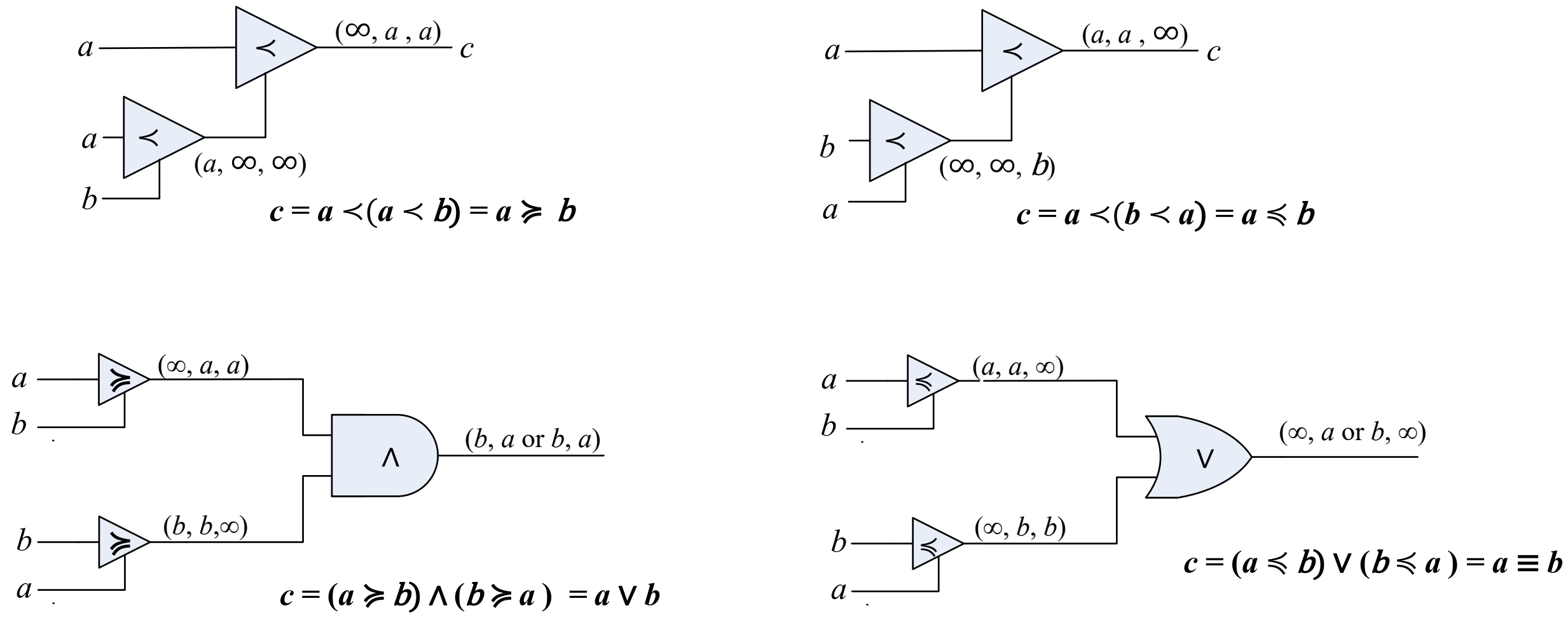

**Figure 3. Examples of functional completeness for primitives +1, $\wedge$, and $\prec$ follow: $\succcurlyeq$ is implemented using only $\prec$. $\preccurlyeq$ is implemented using only $\prec$. $\vee$ is implemented using only $\succcurlyeq$ and $\wedge$. $\equiv$ is implemented using only $\preccurlyeq$ and $\vee$.**

**Theorem:** all the primitive operators in Table 1 can be implemented using only +1, $\wedge$, and $\prec$.

**Proof:** by construction; some are illustrated in Figure 3. All the constructions are given in equations ***20-27***.

### 4.3 Canonical Form

The next objective is to develop a canonical form for *s-t* functions. I.e., a form that represents every *s-t* function uniquely. This is work in progress. Following is a series of steps leading up to a canonical form.

The first step is to partition any *s-t* function into a set of delay functions applied to primary inputs, and a "delay free" function. See Figure 4.

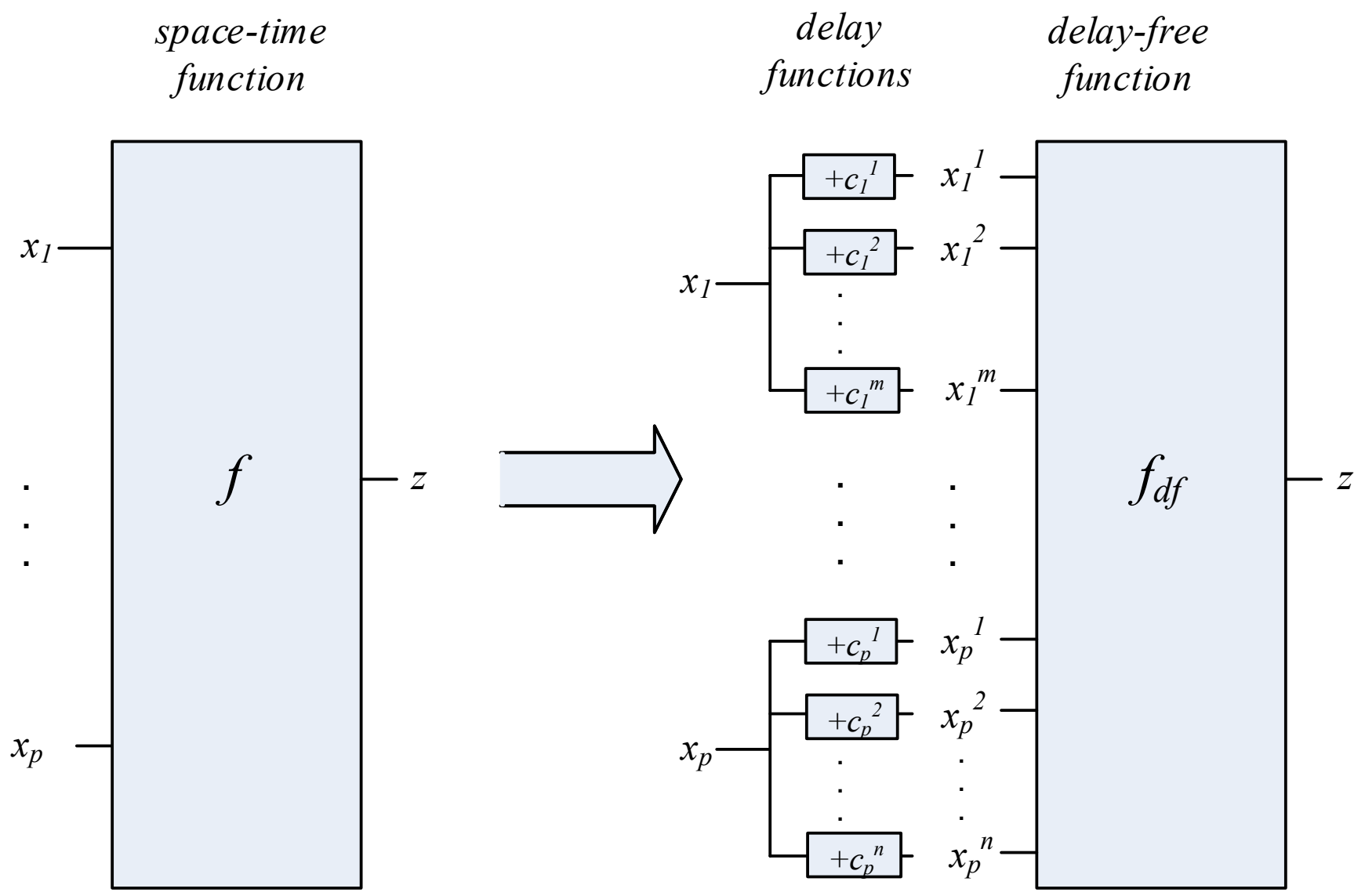

**Figure 4. An *s-t* function decomposed into a set of delay-only functions followed by a delay-free function.**

Any arbitrary feedforward interconnection of *s-t* operators can be decomposed in this fashion by first converting the given network into a fanout-free form by replicating at fanout points and applying invariance repeatedly in order to push all delays to the primary inputs.

Note that after delay functions are pushed to the inputs, the number of inputs applied to the delay-free function $f_{df}$ (marked with superscripts in the figure) will typically be greater than the number of primary inputs for the overall function $f$.

After decomposition is performed, attention can be focused on delay-free functions. For the set of delay-free functions, a *Sequence Table* (analogous to a Truth Table) is constructed and then a canonical form for delay-free *s-t* functions is derived directly from the Sequence Table.

***Sequences***

For a given function evaluation, input variables are assigned specific values which represent the times of temporal events. These specific times (values) can be placed in an ordered acyclic chain, using only the < and = relationships on the *s-t* lattice. This ordered chain is a *sequence*.

All valid sequences for three inputs are illustrated in Table 2. In some cases, a sequence can be expressed more than one way. Note that some sets of temporal relationships do not lead to valid sequences. For example, $a < b$, $b < c$, $c < a$ is cyclic and therefore cannot be described with a valid sequence.

**Table 2. All valid sequences of three variables.**

| *sequence* | *equivalent* |
|---|---|
| $a < b < c$ | |
| $a < b = c$ | $a < c = b$ |
| $a = b < c$ | $b = a < c$ |
| $a = b = c$ | *all permutations* |
| $a = c < b$ | $c = a < b$ |
| $a < c < b$ | |
| $b < a < c$ | |
| $b < a = c$ | $b < c = a$ |
| $b < c < a$ | |
| $b = c < a$ | $c = b < a$ |
| $c < a < b$ | |
| $c < a = b$ | $c < b = a$ |
| $c < b < a$ | |

A delay-free *s-t* function can be defined with a *Sequence Table* (analogous to a Truth Table) that contains a row for every valid sequence. An arbitrary function is defined by assigning an output value to each sequence table row ($f_{example}$, in Table 3). Because the function is delay-free, every non-∞ output value must equal one of the input values. Each row with a non-∞ output corresponds to a *minterm*. An output of ∞ indicates that the corresponding minterm is not present (analogous to a 0 output in a Boolean truth table). Minterms may be specified using an easy-to-read notational shorthand where braces enclose temporal values that are ordered from first to last. Expansion from the shorthand into minterm form is done as:

$$\{ a \bowtie_1 b \bowtie_2 c \} \leftrightarrow a \bowtie_1 b \vee b \bowtie_2 c \text{ where } \bowtie \text{ is either } \prec \text{ or } \equiv.$$

In the example, the minterm denoted as $\{ c \equiv b \prec a \prec \infty \}$ specifies that $a$ and $\infty$ are ordered because the function evaluates to $a$ only if there eventually is a spike on input $a$. This minterm expands to $(c \equiv b) \vee (b \prec a) \vee a$ because $a = a \prec \infty$.

**Table 3 Example sequence table for a function with three input variables.**

| *sequence* | $f_{example}$ | *minterm* |
|---|---|---|
| $a < b < c$ | $b$ | $\{ a \prec b \prec c \}$ |
| $a < b = c$ | ∞ | |
| $a = b < c$ | ∞ | |
| $a = b = c$ | ∞ | |
| $a = c < b$ | ∞ | |
| $a < c < b$ | ∞ | |
| $b < a < c$ | ∞ | |
| $b < a = c$ | $c$ | $\{ b \prec a \equiv c \}$ |
| $b < c < a$ | ∞ | |
| $b = c < a$ | $a$ | $\{ b \equiv c \prec a \prec \infty \}$ |
| $c < a < b$ | ∞ | |
| $c < a = b$ | $a$ | $\{ c \prec a \equiv b \}$ |
| $c < b < a$ | ∞ | |

Before proceeding, keep in mind that *s-t* functions must be causal. Hence, the assignment of output values cannot be completely arbitrary. Some assignments will violate causality; these assignments therefore do not define a valid *s-t* function. This is unlike the Boolean case where every truth table assignment yields a valid Boolean function.

A necessary condition for causality is that any valid sequence must evaluate to either the last or next-to-last variable in the sequence. For example, the sequence $a < b < c$ cannot evaluate to $a$ without violating causality because causality requires that only information available up to time $a$ can be used for determining the output value. At the time event $a$ occurs, it can not be determined whether $b < c$. I.e., one would have to see into the future to determine the eventual relationship between $b$ and $c$.

***Delay-Free Canonical Form***

A delay-free function's canonical form can be generated from the sequence table (analogous to the way a Boolean minterm canonical form can be generated from a truth table).

The canonical form is composed of minterm implicants as illustrated in Table 3. The overall function output is the *min* of the minterms. Consequently:

$$f_{example} = \{c \equiv b \prec a \prec \infty\} \wedge \{b \prec a \equiv c\} \wedge \{c \prec a \equiv b\} \wedge \{a \prec b \prec c\}$$

$$= [(c \equiv b) \vee (b \prec a) \vee a] \wedge [(b \prec a) \vee a \equiv c)] \wedge [(c \prec a) \vee (a \equiv b)] \wedge [(a \prec b) \vee (b \prec c)]$$

***General Canonical Form***

The above describes a delay-free canonical form. Adding input delays does not necessarily produce a canonical form for the overall network. At least it remains to be shown.

Also, when recombined with the delay network, some simplifications can be performed. That is, there may be minterms that contain relations involving the same input variable with different delays, e.g., $a + 3 \prec a + 7$ or $a + 7 \prec a + 3$. In the first case, this relation can simply be removed from the minterm because it is always satisfied. In the second case, the entire minterm should be removed because it is never satisfied.

## 4.4 Standard Form

Minterms can be generalized to implicants, and a *standard form* is composed of implicants (Figure 5). A *general sequence* is not required to contain all input variables as in the case of a minterm sequence. The only requirement is that there are no contradictory relations in the implicant.

For example, $(a \prec b) \vee (b \equiv d)$ is an implicant for a function with four variables. Another example is $(a \prec b) \vee (c \prec d)$. The first example can be written as a single sequence $a < b \equiv d$. The second example can only be written as two subsequences. A general sequence, then, is analogous to a Boolean implicant, and a function can be specified as a set of general sequences (subject to causality).

It is sometimes convenient to add the $\preccurlyeq$ operator to standard form sequences. In sequence notation: $\{ a \bowtie_1 b \prec c \bowtie_3 d\} \wedge \{ a \bowtie_1 b \equiv c \bowtie_3 d \} \leftrightarrow \{ a \bowtie_1 b \preccurlyeq c \bowtie_3 d \}$ , where $\bowtie_i$ is $\prec$, $\preccurlyeq$, or $\equiv$.

*proof:*

| | |
|---|---|
| $\{ a \bowtie_1 b \prec c \bowtie_3 d\} \wedge \{ a \bowtie_1 b \equiv c \bowtie_3 d \} \leftrightarrow [\, a\bowtie_1 b \vee b \prec c \vee c\bowtie_3 d \,] \wedge [\, a\bowtie_1 b \vee b \equiv c \vee c\bowtie_3 d \,]$ | |
| $= [\, a\bowtie_1 b \vee c\bowtie_3 d \,] \vee [\, b \prec c \wedge b \equiv c \,]$ | distributive |
| $= [\, a\bowtie_1 b \vee c\bowtie_3 d \,] \vee [\, b \preccurlyeq c \,]$ | by defn. |
| $= [\, a\bowtie_1 b \vee b \preccurlyeq c \vee c\bowtie_3 d \,] \leftrightarrow \{ a \bowtie_1 b \preccurlyeq c \bowtie_3 d \}$ | associative, commutative |

In the standard form, primary inputs are fanned out and delayed by function-dependent amounts. Then, the delayed inputs feed a single level of relational operators $\prec$, $\equiv$, and $\preccurlyeq$. The outputs of the relational operators feed a set of ∨ (*max*) operators. Each ∨ operator implements a temporal implicant. Then, all the ∨ operators feed a ∧ (*min*) operator that produces the function's output.

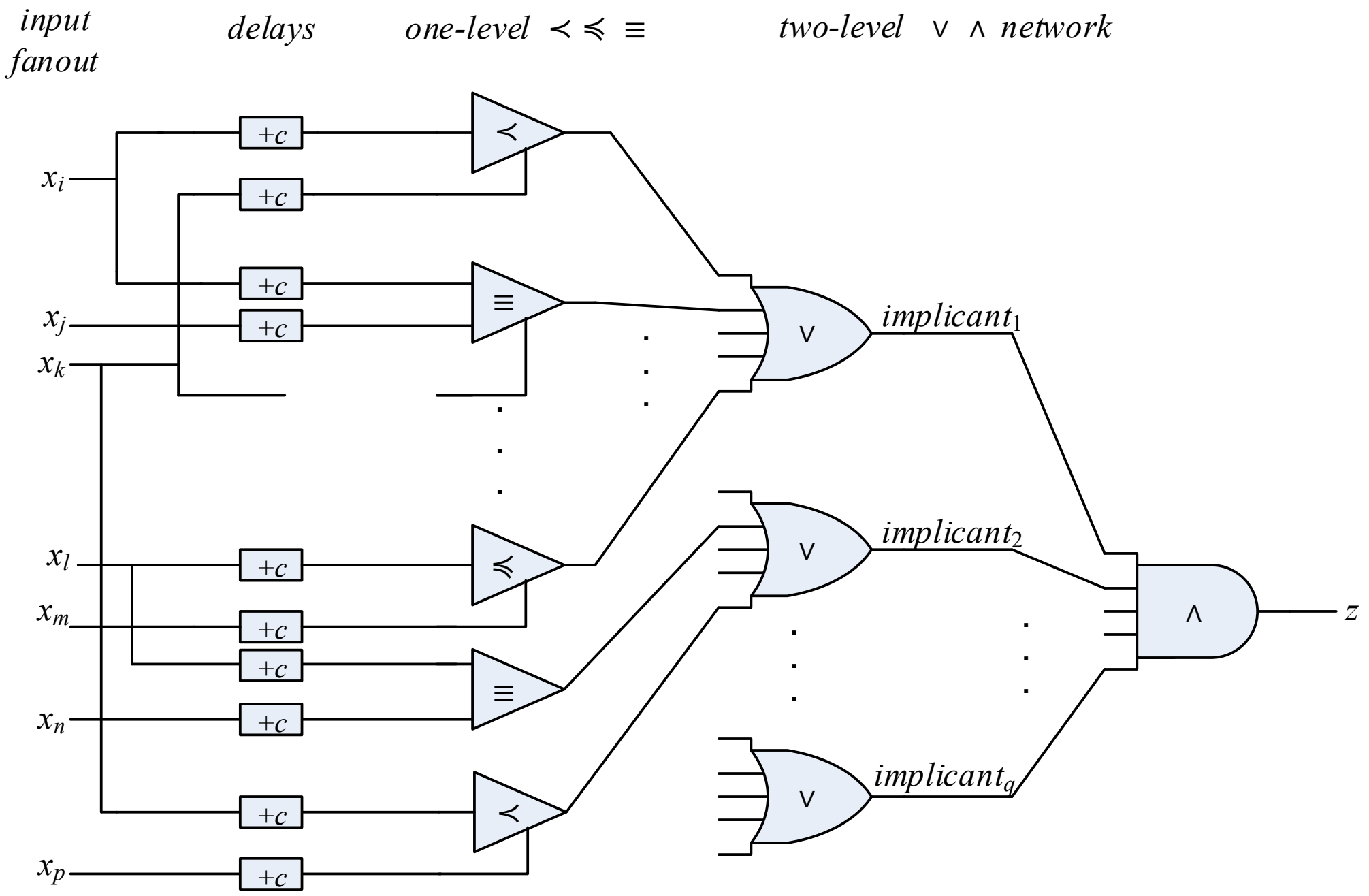

**Figure 5. Standard form: all feedforward *s-t* networks can be reduced to this form.**

### *Conversion to Standard Form*

Any arbitrary feedforward network composed of pairwise primitive operators can be reduced to the standard form. This process is analogous to converting an arbitrary multi-level feedforward Boolean logic network into a sum-of-products form. One approach for arriving at a standard form is:

1) Reduce all 2-ary operators to +1, ≺ , ≼, ≡, ∧, and ∨.

2) If a network contains internal fanout, use subnetwork replication to expand it to a network with fanout only at primary inputs.

3) Push all delays ($+c$) back to primary inputs using invariance, e.g., ***17***, ***18***, ***19.*** Combine serial delays to form a single, larger delay.

4) Use the distributive properties repeatedly to push the ≺, ≼, and ≡ operators back to the delayed inputs. This is roughly analogous to using deMorgan's law to push *not* gates to the primary inputs in Boolean algebra sum-of-products form, albeit more complicated.

5) After step 4) the network will consist of the following layers: 1) fanned-out primary inputs feeding delays, 2) a layer of relational operators ≺, ≼, and ≡, 3) a multi-level network consisting only of ∧, and ∨.

6) The ∧, ∨ network can then be reduced to standard form using the associative, commutative, and distributive properties.

The final network consists of fanned-out primary inputs passing through a layer of delays, a layer of relational operators, and a two-level ∨, ∧ network as in Figure 5. Each implicant (∨ term) specifies a set of timing relations that must hold simultaneously for the implicant to be satisfied. That is, an implicant

“implies” that the output is non-$\infty$. The $\wedge$ operation forming the final output is satisfied if any of the implicants is satisfied.

At this point it appears that a useful set of operators for design are: $\wedge$, $\vee$, $\prec$, $\equiv$, $\preccurlyeq$, +1 as used in the canonical and standard forms. One can express functions in tabular sequence table form based on these operators.

## 4.5 Completeness

Any Boolean function can be implemented by composing *and*, *or*, and *not* gates; hence, these three primitives are functionally complete. A *nand* or *nor* gate alone is also functionally complete.

For *s-t* computing networks, we might like to identify sets of primitives that are functionally complete. Given the sequence of transformations from arbitrary form to standard form to canonical form to the sequence table, it follows that +1, $\prec$ *, and* $\wedge$ *are complete for all s-t functions that can be implemented as an arbitrary feedforward interconnection of s-t operators.*

It has been shown thus far that 1) An arbitrary network can be decomposed into a set of delays and a delay-free function. 2) There is a canonical form for the complete set of delay-free functions. This form uses five operators, but these can be reduced to the three +1, $\prec$, and $\wedge$. Hence, completeness follows for all *s-t* functions that can be implemented as a feedforward interconnection of *s-t* operators.

It is important to observe, however, that the decomposition of an arbitrary network into delays and delay-free portions starts with an implementation wherein delays are always implemented as explicit operators. To extend to all functions, with no restriction on delay implementation, the delay decomposition must apply *independently* of implementation. Although it makes intuitive sense that this can be done, it remains to be shown.

***The xor/xnor Issue***

An important piece of artificial neural network (ANN) lore is that after the first weighted artificial neurons were proposed -- *perceptrons* -- Minsky and Papert [10] showed that a perceptron can not implement all Boolean functions. In particular, *xor* and its complement, *xnor* (or *equivalence*), can not be implemented. Hence, perceptrons were generally considered to be deficient, and, according to the lore, this led to a long connectionist “winter”. The apparent shortcoming was eventually resolved via multiple layer perceptrons, and artificial neural networks re-emerged as an active research topic in the late 1980s.

Given its significance in the development of ANNs, consider the *xor* problem in the context of *s-t* functions. The issue to be addressed is whether *xor*, or *xor* -like, functions can be implemented using *s-t* operations. The conventional *xor logic* function is defined over a lattice using logical operators. In contrast, the *s-t* algebra is defined over a lattice restricted to operations consistent with the flow of time: causality and invariance.

Consider the *s-t* algebra with elements $N_0^\infty$ and a set of causal and invariant operators. Although the following discussion holds for all points in the algebra, for the sake of argument, consider only 0 and $\infty$. That is, “binarize” values by mapping any spike to a spike at time = 0, and no spike to time = $\infty$. Then, a temporal *xor* function is given in Table 4.

**Table 4. Temporal *xor***

| $x_1$ | $x_2$ | $z$ |
|---|---|---|
| 0 | 0 | ∞ |
| 0 | ∞ | 0 |
| ∞ | 0 | 0 |
| ∞ | ∞ | ∞ |

The temporal *xor* function is both invariant and causal and is therefore an *s-t* function. On the other hand, the temporal *xnor* (Table 5) is not. In the bottom row, there is an output spike at $t = 0$ if neither of the inputs has a spike at any time in the future. Because it is not causal, there is no *s-t* network that will implement the *xnor* function.

**Table 5. Temporal *xnor***

| $x_1$ | $x_2$ | $z$ |
|---|---|---|
| 0 | 0 | 0 |
| 0 | ∞ | ∞ |
| ∞ | 0 | ∞ |
| ∞ | ∞ | 0 |

Consequently, although there is no *s-t xnor*-like operation, there is a temporal *xor-like* operation. This is certainly better than not being able to implement either, and a temporal *xor* alone may be adequate for supporting useful *s-t* functions, especially the neocortical ones.

In general, for a function to be causal, the all-∞ input must always yield an ∞ output. This property holds for the *xor* but not the *xnor*. In Boolean algebra, the *xnor* is an *xor* with an inverter. In the *s-t* algebra there is no inverter, because inversion is tantamount to going backward in time.

## 5. Temporal Neural Networks

Temporal Neural Networks (TNNs)[3] perform communication and computation using spikes that model the voltage pulses, or action potentials, that are present in biological neural networks [6][9][12][13][14]. In model TNNs, as considered here, a vector of information is conveyed on multiple lines with at most one spike per line. Values are encoded based on timing relationships among individual spikes [5].

Most proposed TNN models are feedforward, spike-based computing networks that satisfy the constraints of causality and invariance. To emphasize this point, in this section some commonly used TNN components are constructed using only the *s-t* primitives. The following TNN elements considered: 1) SRM0 neurons with arbitrary response functions, 2) Synaptic weight interfacing via micro-weight networks, 3) Winner-take all (WTA) inhibition.

These example implementations serve two purposes: to illustrate *s-t* computation as a general proposition and to provide a basis for TNN simulation and/or direct hardware implementation.

### 5.1 SRM0 Neurons

Excitatory neurons employ an SRM0 model [7]; see Figure 6. In the SRM0 model, input $x_i$ connects to the neuron body via synapse *i* having weight $w_i$. 1) if there is a spike on $x_i$, then the value (time) of the associated synaptic weight $w_i$ selects a pre-defined *response function*; 2) in the neuron body, the synaptic response functions are linearly *summed* (*integrated*), yielding a net *body potential*; 3) when, and if, the body potential reaches a *threshold* value θ, a spike on output *z* is produced at that time.

Typically, different weights map to response functions that differ only in amplitude, with a high weight indicating a high amplitude. However, in general this is not a requirement; weights can be assigned arbitrary response functions.

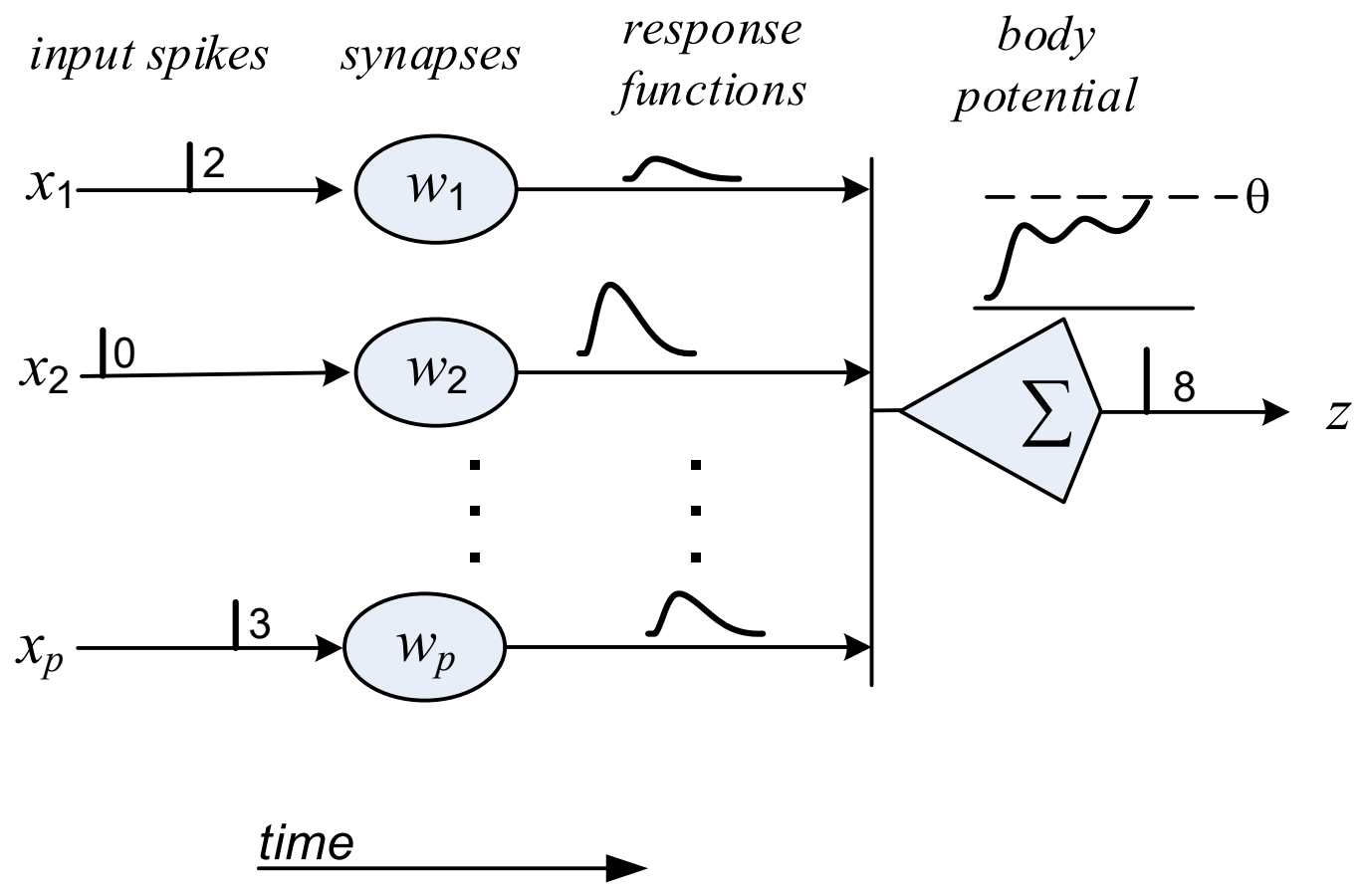

**Figure 6. SRM0 neuron model. Spikes denote points in time. In the figure, values are shown next to their associated spikes. The first spike in time assigned the value 0, and all other spikes encode values with times relative to the first.**

#### 5.1.1 Response Functions

A *response function* ρ(*w*, *t*) maps the non-negative integers representing a synaptic weight and discrete time, respectively, onto integers representing a neuron's body potential. The only constraints are that the function's range is bounded and for a given value of *w*, the function's output remains constant after some time $t_{max}$. This very broad response function definition includes discretized versions of all proposed response functions of which the author is aware.

[3] Maass [8] applied the term "Spiking Neuron Network" (SNN) to a class of networks that use spike times to encode information (as TNNs do). However, over time, the term "SNN" has broadened to include networks that do not use individual spike time relationships to encode information; some use spike rates, instead.

As an example, Figure 7 is a discretized low resolution version of the commonly used biexponential response function. For this response function, $w = 5$, which, in this example is the maximum amplitude of the response function.

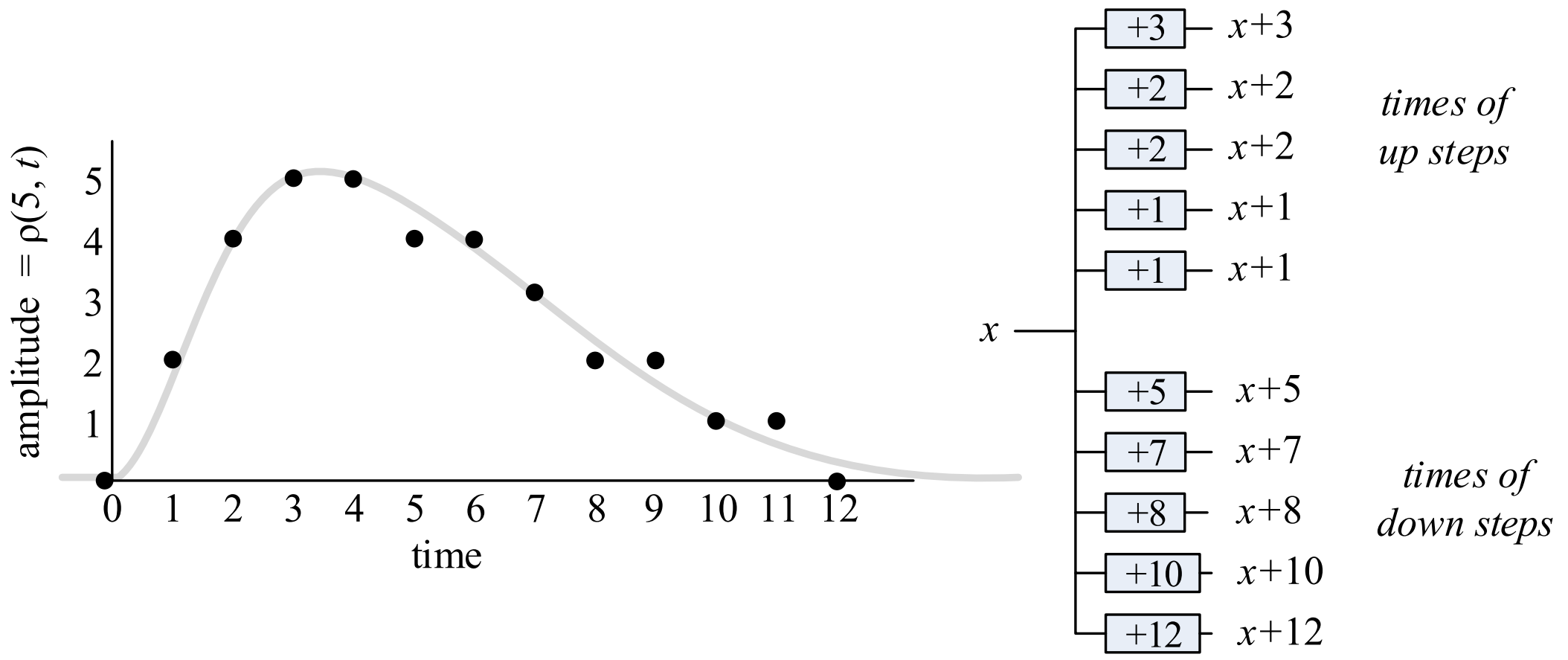

**Figure 7. Biexponential response function and corresponding *s-t* fanout network.**

First consider modeling the response function for a single input, *x*. Due to its discrete nature, the response function's amplitude at any given time can be expressed as an integer number of *amplitude units* (either positive or negative). By sequencing through values of *t*, beginning with $t = 0$ a response function can be composed of a sequence of *up steps* and *down steps*. Each "step" is a single discrete amplitude unit.

Accordingly, to construct the response function for input *x*, the input *x* is fanned out, with increment values placed on the fanned-out lines. The fanouts are divided into two groups: *up* fanouts and *down* fanouts. For each $t \geq 1$, define $s = \rho(w, t) - \rho(w, t\text{-}1)$; for the end case $t = 0$, define $s = \rho(w, 0)$. If *s* is positive, then *x* fans out *s* times to the *up* fanout network; each of the fanouts is assigned an associated increment value of *t*. If *s* is negative, then fanout *s* times to the *down* fanout network, each of the fanouts is assigned an associated value of *t*.

Figure 12 (right) illustrates the fanout/increment network for a biexponential response function. The function takes two *up* steps at $t = 1$, two more *up* steps at $t = 2$, a single *up* step at $t = 5$, then a series of *down* steps at $t = 5, 7, 8, 10, 12$.

An *s-t* implementation of a neuron body takes multiple response functions in the *up*/*down* step format as inputs, integrates them, and emits an output spike at the time the threshold θ is reached. The sort function is a key building block for a neuron body implementation.

### 5.1.2 Bitonic Sorting Networks

*Sort* is an important *s-t* function that has many potential applications. First observe that *sort* is causal and invariant. When numbers are sorted from smallest to largest, the position of any given number in the sorted list depends only on the locations of smaller or equal numbers. Any larger numbers are irrelevant in determining its location in the sorted list. *Sort* is invariant because adding a constant to all the inputs does not change the sorted order, and the sorted outputs have the same added constant.

A bitonic sorting network [2] is constructed using two-input, two-output compare elements, each consisting of a *min* function and a *max* function (see 8).

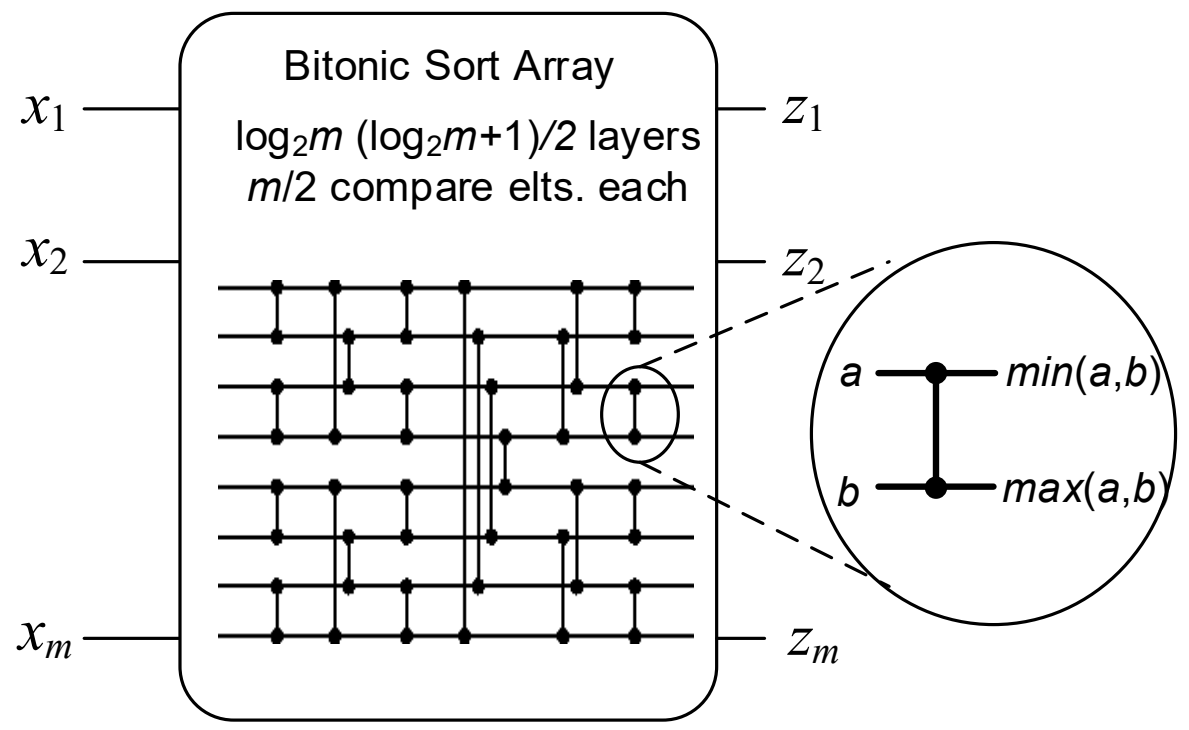

**8. A bitonic sorting network consists of interconnected *min*/*max* comparator elements.**

Because it uses only *min* and *max* elements, which are both causal and invariant, the entire sorting network must be causal and invariant (Lemma 1).

### 5.1.3 Neuron Body Implementation

Given input response functions implemented as *s-t* fanout networks, the remainder of the SRM0 neuron implementation is constructed as illustrated in Figure 9.

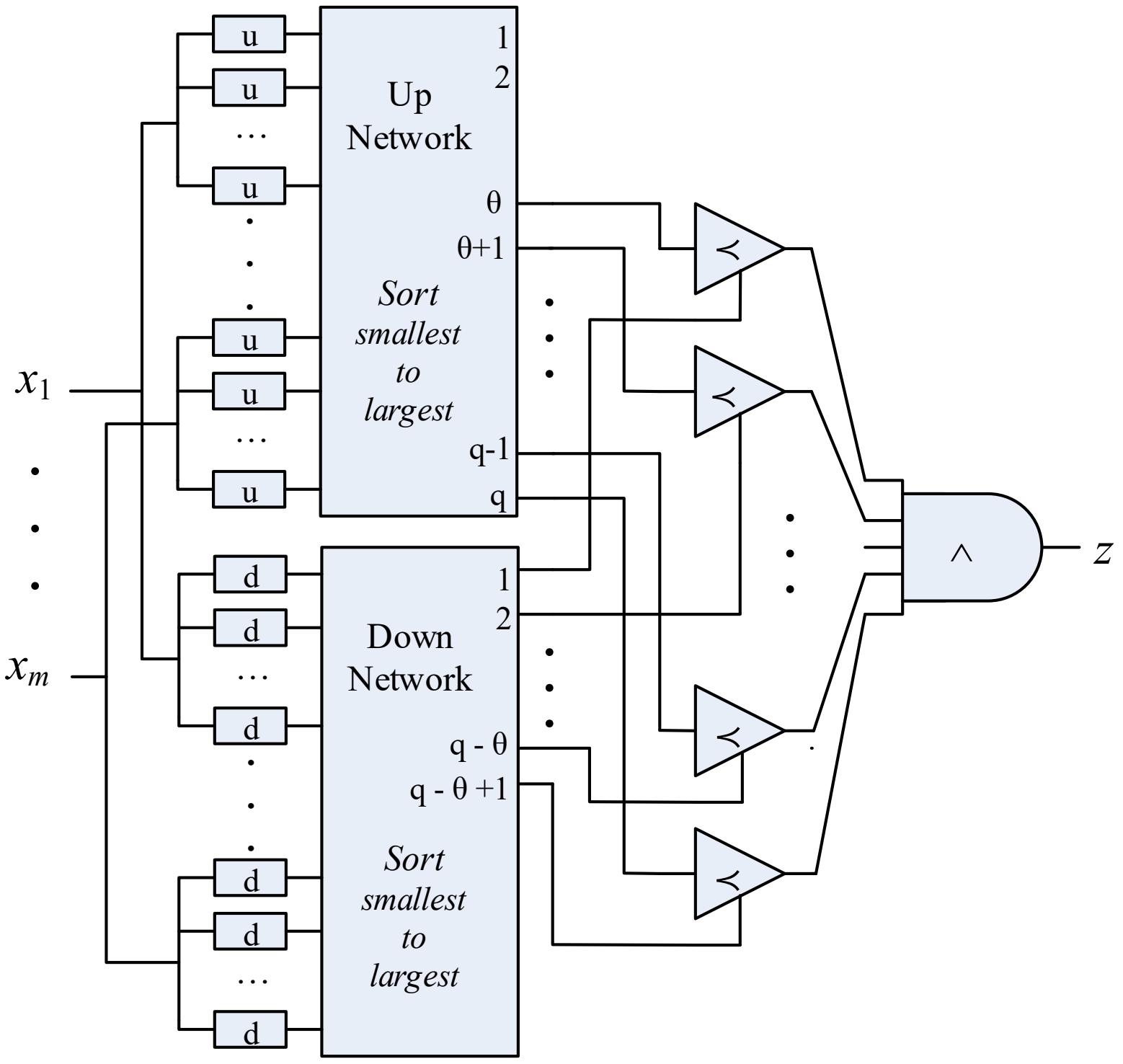

**Figure 9. SRM0 implementation using *s-t* primitives.**

The fanned-out *up*/*down* increment values are the inputs to two *sort* networks (inputs are denoted generically in the figure as “u” and “d”) for all $x_i$. Collectively, these specify the response behavior for all the primary inputs. The *up* steps are sent to one sort network, and *down* steps are sent to the other. After sorting, the two sets of sort network outputs are combined via a series of ≺ blocks. The objective is to determine the first time (value) for which the number of *up* steps exceeds the number of *down* steps by the threshold value θ. That is, the *inhibit* blocks determine if the time of the θ + *i*th *up* step occurs before the *i*th *down* step. If so, the non-∞ value of the θ + *i*th *up* step is the input to the final *min* function.

Otherwise, the input value to the final *min* function is ∞. The *min* function's output is therefore the first time the number of *up* steps exceeds the number of *down* steps by the amount θ. This is the first time the threshold is crossed.

## 5.2 Implementing Synaptic Weights

In a typical neuron model, synaptic weights determine response functions, which in turn determine an individual neuron's function, and, consequently, the overall neural network's function. In neural networks, synaptic weights are typically established via a process where inputs from a training set are applied, outputs are observed, and weights are adjusted to reflect patterns inherent in the training inputs. After training, the learned synaptic weights become part of a neuron's function definition. In a sense, an untrained network is configured, or programmed, depending on the weight settings.

Training processes are not considered here, because, among other things, the training process is typically not an *s-t* function. In general, the state update functions may not be causal and invariant (although some portions of them may be). It is assumed here that weights and weight updates are implemented as a classical state machine, with the weights being explicit binary state.

Even though the weights are explicit binary state, they must interface with the *s-t* network in order to participate in response function generation. A primitive interface mechanism is conceptually an enable/disable switch (see Figure 10). In Figure 10, a binary *micro-weight* $\mu$ input to an ≺ function maps either to 0 or ∞ prior to *s-t* computation. If input $\mu = 0$, then the output of its associated ≺ operator is ∞, regardless of the input $x$. If input $\mu = 1$, then the $x$ input value passes through to the output.

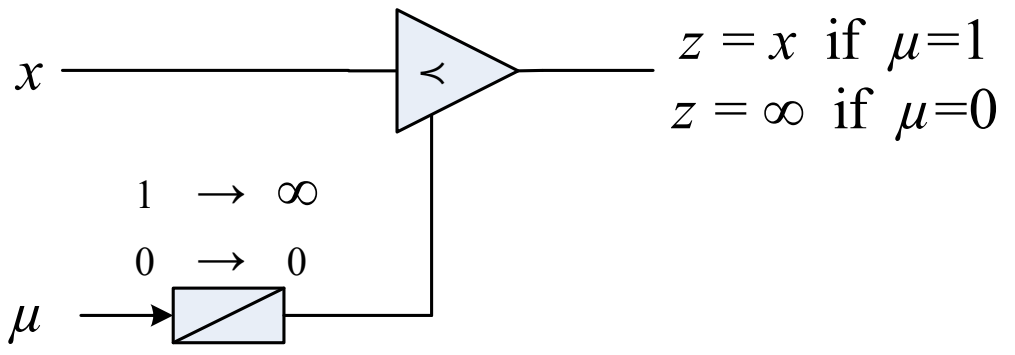

**Figure 10. A binary micro-weight $\mu$ maps to a spike at time 0 (if $\mu$ = 0) or ∞ (if $\mu$ = 1). These spikes then enable, or inhibit, the passage of input x by using the ≺ function.**

Given the interface just described, one can design a fanout/increment network that takes binary-coded weights as inputs and maps them to micro-weight settings that control delay operators. The selected delay operators then define a response function in *up*/*down* form.

The structure of the network depends on the structure of the response functions (and the designer's ingenuity). As a simple example, the network shown in Figure 11 implements a set of four response functions, having different amplitudes corresponding to a range of synaptic weights. The synaptic weights are set via four micro-weights $\mu_1$ through $\mu_4$. In the example, the weight is encoded in a thermometer code so that the bits of the code directly map to the $\mu_i$. If the synaptic weight to 3, for example, then the micro-weights form the vector [1, 1, 1, 0].

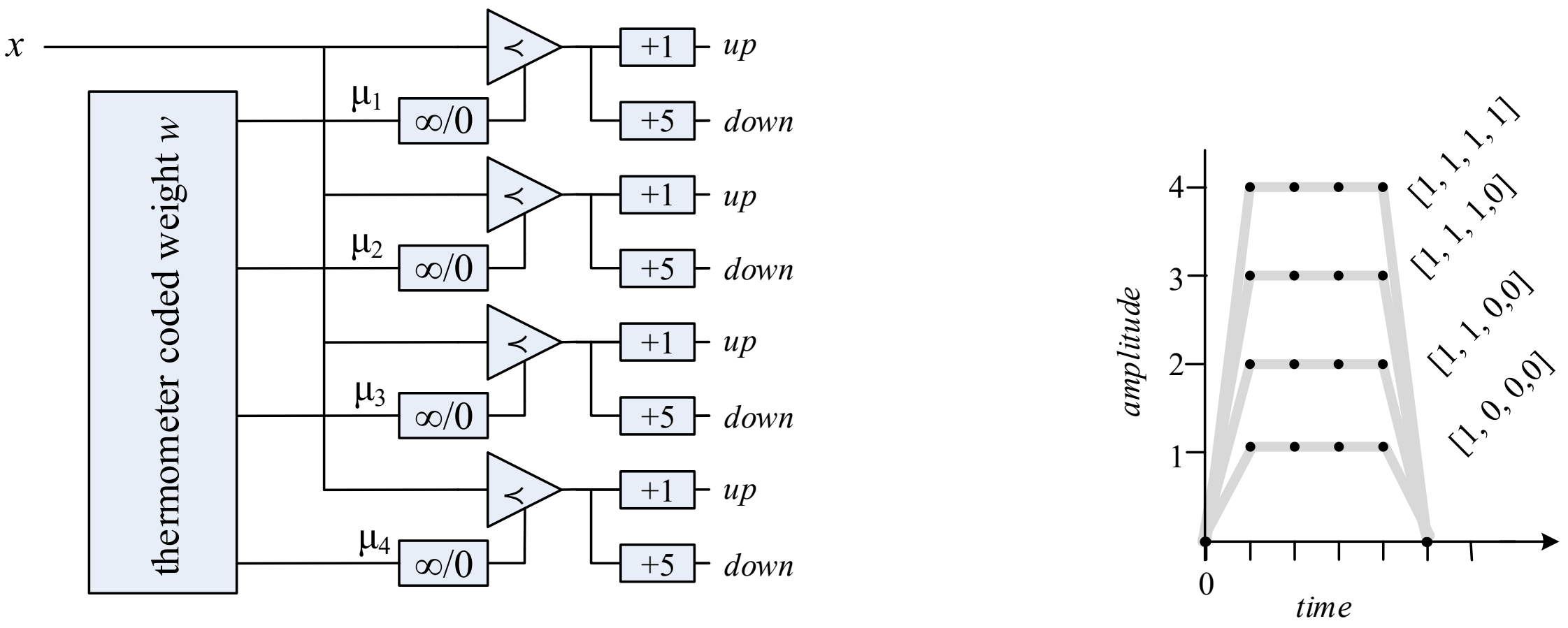

**Figure 11. Modeling response functions for a range of synaptic weights. In this example, the range is 0 to 4. Synaptic weights are determined via a vector of micro-weights [ $\mu_1$ , $\mu_2$ , $\mu_3$ , $\mu_4$ ].**

## 5.3 Inhibition: Winner-Take-All (WTA) Networks

In the neuroscience literature, inhibition is typically modeled as WTA lateral inhibition. In the case of TNNs, the "winners" are the first spikes in a volley, so winner-take-all inhibits all but the first spikes. In general, what is meant by "first" may be parameterized. It may the first *k* spikes, or the spikes that appear within some time window beginning with the first, or some combination.

Figure 12 is the implementation of a simple 1-WTA network where only the spikes occurring at relative time 0, are allowed to pass; all the others are inhibited. The $\wedge$ gate finds the time of the first spike(s), and that time, delayed by 1 time unit, inhibits all the others. In this implementation, if there is a tie for "first", all the first spikes are allowed to pass. By adding additional circuitry, more elaborate methods for selecting certain tying output spikes can be implemented.

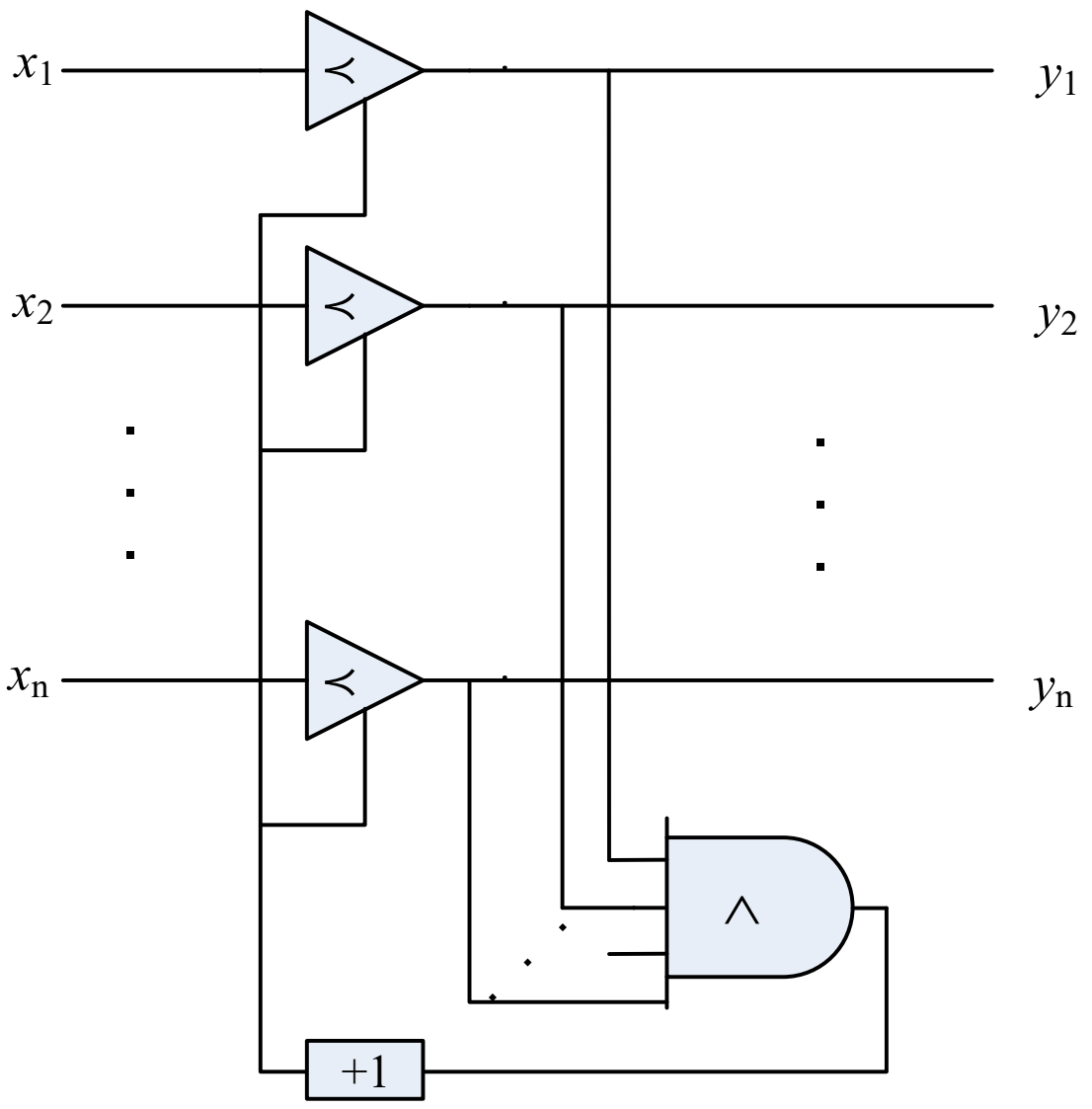

**Figure 12. Winner-take-all network. Only the first spike(s) pass through uninhibited.**

## 6. Relationship to Allen's Interval Algebra

Allen's algebra [1] operates on relationships involving time intervals. Figure 13 contains the 13 basic relations (6 are non-commutative and therefore yield two separate relations depending on the order of the input variables). These basic relations may be composed to form more complex indefinite intervals that underpin algorithmic reasoning about temporal relationships.

| Relation | Illustration | Interpretation |
|---|---|---|
| $X < Y$<br>$Y > X$ | X<br>Y | X takes place before Y |
| $X \mathbf{m} Y$<br>$Y \mathbf{mi} X$ | X<br>Y | X meets Y (*i* stands for ***inverse***) |
| $X \mathbf{o} Y$<br>$Y \mathbf{oi} X$ | X<br>Y | X overlaps with Y |
| $X \mathbf{s} Y$<br>$Y \mathbf{si} X$ | X<br>Y | X starts Y |
| $X \mathbf{d} Y$<br>$Y \mathbf{di} X$ | X<br>Y | X during Y |
| $X \mathbf{f} Y$<br>$Y \mathbf{fi} X$ | X<br>Y | X finishes Y |
| $X = Y$ | X<br>Y | X is equal to Y |

**Figure 13. Operations in Allen's Interval Algebra.** (from Tales Paiva Nogueira via ResearchGate)

In terms of the *s-t* algebra, Allen's basic intervals can be specified by events that identify their starting and finishing times. For the interval *X* the events are $X_s$ ("*X* starts") and $X_f$ ("*X* finishes"). Therefore, an implied relation that always holds is $X_s \prec X_f$ .

The basic interval relations of Allen's algebra can be expressed in the *s-t* algebra as:

*X* takes place before *Y*: $X_f \prec Y_s$

*X* meets *Y*: $X_f \equiv Y_s$

*X* overlaps with *Y*: $X_s \prec Y_s \vee Y_s \prec X_f \vee X_f \prec Y_f$

*X* starts *Y*: $X_s \equiv Y_s \vee X_f \prec Y_f$

*X* during *Y*: $Y_s \prec X_s \vee X_f \prec Y_f$

*X* finishes *Y*: $Y_s \prec X_s \vee X_f \equiv Y_f$

*X* is equal to *Y*: $X_s \equiv Y_s \vee X_f \equiv Y_f$

***Example from Wikipedia:***

*During dinner, Peter reads the newspaper. Afterwards, he goes to bed.*

Events:

$D_s$ == dinner starts; $D_f$ == dinner finishes

$R_s$ == Peter starts reading paper; $R_f$ == Peter finishes reading paper.

$B_s$ == Peter goes to bed.

Then the following *s-t* expression describes the system:

$D_s \prec R_s \vee R_f \prec D_f \vee D_f \prec B_s$

Dinner starts before reading starts and reading is finished before dinner is finished. Furthermore, dinner is finished before going to bed.

This expression corresponds to the sequence: $D_s < R_s < R_f < D_f < B_s$ .

***Realtime Characteristics of s-t Algebra***

An expression in Allen's algebra is generally not used for specifying a computation that takes actual input values and produces an output value. Rather, Allen's algebra is typically used as an *analysis* tool wherein a collection of relations form the basis for analyzing timing relationships. Two primary analysis problems follow.

The first problem is to determine the *strongest implied relation* between two intervals $X$ and $Y$. This is done by finding all chains of inference between $X$ and $Y$ and then taking their intersection. Each chain of inference constrains the relationship, so the intersection is a constraint (the strongest implied relation) that covers all possible chains between $X$ and $Y$.

The *satisfaction* problem is to determine for a collection of relations whether there is any set of intervals such that all the relations in the collection are true. The satisfaction problem is known to be NP complete.

In the *s-t* algebra, an implicant is a chain of inferences combined via the *max* ($\vee$) operator. A collection of inferences is formed by taking the *min* ($\wedge$) of the set of implicants.

An *s-t* expression is useful for both *analysis* and *synthesis* -- i.e., for designing temporal computing devices.

***Analysis***

*Satisfaction*: determining whether a given network has at least one set of inputs that yield a non-$\infty$ output.

*Strongest Implied Relation:* The largest set of input relationships that are the same for all input patterns satisfying the expression.

***Synthesis***

Conceptually, an *s-t* expression describes a computation that can be implemented and evaluated in real time. I.e., an expression being evaluated "observes" events at the time they occur, and then, as early as it can, it determines whether the observed sequence is a satisfactory sequence.

For example, if an implicant is $D_s \prec R_s \vee R_f \prec D_f \vee D_f \prec B_s$ and if input events occur at the times $D_s$ = 7:00 PM, $R_s$ = 7:10, $R_f$ = 8:00, $D_f$ = 8:10, and $B_s$ = 9:00, then the conditions are all satisfied. Further, it is known they have all been satisfied at 8:10, so the output signal occurs at 8:10. (We don't need to know Peter's exact bedtime as long as we know it is after 8:10). However, if one of the conditions fails, say that reading starts before dinner, then the conditions will never be satisfied, and the output is "$\infty$".

Finally, if it is possible that Peter never goes to bed, then we need to add the additional term "$\vee B_s$" to yield $D_s \prec R_s \vee R_f \prec D_f \vee D_f \prec B_s \vee B_s$ . Then, if Peter never goes to bed, the implicant evaluates to $\infty$. If he does go to bed, then the soonest we know the implicant is satisfied is time $B_s$.

Hence, by applying values which represent the times of events, an implicant indicates whether a particular set of values (times) satisfies the ordering relation specified by the implicant, and, if so, the output value is the earliest time at which the conclusion of satisfaction can be reached. If it is not consistent, then the output value is $\infty$ ; i.e., there is never a time when the relation will be satisfied. When given a collection of implicants, and more than one of the time sequences is satisfied, then the output corresponds to the earliest one; i.e., this is the earliest we know that at least one of the sequences is satisfied.